


\font\titlefont = cmr10 scaled\magstep 4
\font\sectionfont = cmr10
\font\littlefont = cmr5

\magnification = 1200

\global\baselineskip = 1.2\baselineskip
\global\parskip = 4pt plus 0.3pt
\global\abovedisplayskip = 18pt plus3pt minus9pt
\global\belowdisplayskip = 18pt plus3pt minus9pt
\global\abovedisplayshortskip = 6pt plus3pt
\global\belowdisplayshortskip = 6pt plus3pt


\def\endignore{}
\def\ignore #1\endignore{}

\newcount\dflag
\dflag = 0


\def\monthname{\ifcase\month
\or Jan \or Feb \or Mar \or Apr \or May \or June%
\or July \or Aug \or Sept \or Oct \or Nov \or Dec
\fi}

\def\timestring{{\count0 = \time%
\divide\count0 by 60%
\count2 = \count0
\count4 = \time%
\multiply\count0 by 60%
\advance\count4 by -\count0
\ifnum\count4 < 10 \toks1 = {0}
\else \toks1 = {} \fi%
\ifnum\count2 < 12 \toks0 = {a.m.}
\else \toks0 = {p.m.}
\advance\count2 by -12%
\fi%
\ifnum\count2 = 0 \count2 = 12 \fi
\number\count2 : \the\toks1 \number\count4%
\thinspace \the\toks0}}



\def\endtitle{}
\def\title#1\endtitle{\vskip.5in\titlefont
\global\baselineskip = 2\baselineskip
#1\vskip.4in
\baselineskip = 0.5\baselineskip\rm}

\def\endauthors{}
\def\authors#1\endauthors{#1}

\def\endabstract{}
\def\abstract#1\endabstract{\vskip .3in%
\centerline{\sectionfont\bf Abstract}%
\vskip .1in
\noindent#1}

\newcount\nsection
\newcount\nsubsection

\def\section#1{\global\advance\nsection by 1
\nsubsection=0
\bigskip\noindent\centerline{\sectionfont \bf \number\nsection.\ #1}
\bigskip\rm\nobreak}

\def\subsection#1{\global\advance\nsubsection by 1
\bigskip\noindent\sectionfont \sl \number\nsection.\number\nsubsection)\
#1\bigskip\rm\nobreak}


\def\appendix#1#2{\bigskip\noindent%
\centerline{\sectionfont \bf Appendix #1.\ #2}
\bigskip\rm\nobreak}


\newcount\nref
\global\nref = 1

\def\ref#1#2{\xdef #1{[\number\nref]}
\ifnum\nref = 1\global\xdef\therefs{\noindent[\number\nref] #2\ }
\else
\global\xdef\oldrefs{\therefs}
\global\xdef\therefs{\oldrefs\vskip.1in\noindent[\number\nref] #2\ }%
\fi%
\global\advance\nref by 1
}

\def\listrefs{\vfill\eject\section{References}\therefs}


\newcount\cflag
\newcount\nequation
\global\nequation = 1
\def\eqlabel{(1)}

\def\nexteqno{\ifnum\cflag = 0
\global\advance\nequation by 1
\fi
\global\cflag = 0
\xdef\eqlabel{(\number\nequation)}}

\def\lasteqno{\global\advance\nequation by -1
\xdef\eqlabel{(\number\nequation)}}

\def\label#1{\xdef #1{(\number\nequation)}
\ifnum\dflag = 1
{\escapechar = -1
\xdef\draftname{\littlefont\string#1}}
\fi}

\def\clabel#1#2{\xdef\eqlabel{(\number\nequation #2)}
\global\cflag = 1
\xdef #1{\eqlabel}
\ifnum\dflag = 1
{\escapechar = -1
\xdef\draftname{\string#1}}
\fi}

\def\cclabel#1#2{\xdef\eqlabel{#2)}
\global\cflag = 1
\xdef #1{\eqlabel}
\ifnum\dflag = 1
{\escapechar = -1
\xdef\draftname{\string#1}}
\fi}


\def\eeq{}

\def\eqnn #1\eeq{$$ #1 $$}

\def\eq #1\eeq{\xdef\draftname{\ }
$$ #1
\eqno{\eqlabel \rlap{\ \draftname}} $$
\nexteqno}



\def\eol{& \eqlabel \rlap{\ \draftname} \crcr
\nexteqno
\xdef\draftname{\ }}

\def\eeol{& \eqlabel \rlap{\ \draftname}
\nexteqno
\xdef\draftname{\ }}

\def\eolnn{\cr
\global\cflag = 0
\xdef\draftname{\ }}


\def\eqa #1\eeq{\xdef\draftname{\ }
$$ \eqalignno{ #1 } $$
\global\cflag = 0}


\def\etal{{\it et.al.\/}}

\def\via{{\it via\/}}


\def\jetp#1#2#3{{\it JETP Lett.} {\bf #1} (19#2) #3}

\def\npb#1#2#3{{\it Nucl. Phys.} {\bf B#1} (19#2) #3}
\def\plb#1#2#3{{\it Phys. Lett.} {\bf #1B} (19#2) #3}

\def\prd#1#2#3{{\it Phys. Rev.} {\bf D#1} (19#2) #3}

\def\prl#1#2#3{{\it Phys. Rev. Lett.} {\bf #1} (19#2) #3}

\def\sjnp#1#2#3#4#5#6{{\it Yad. Fiz.} {\bf #1} (19#2) #3
[{\it Sov. J. Nucl. Phys.} {\bf #4} (19#5) #6]}
\def\zpc#1#2#3{{\it Zeit. Phys.} {\bf C#1} (19#2) #3}


\global\nulldelimiterspace = 0pt



\def\frac#1#2{{{#1} \over {#2}}\,}  

\def\nth#1{{1\over #1}}


\def\Dsl{\hbox{/\kern-.6000em\it D}} 
\def\dsl{\hbox{/\kern-.5600em$\partial$}}
\def\pxpsl{\hbox{/\kern-.5600em$p$}}
\def\ssl{\hbox{/\kern-.5600em$s$}}
\def\epssl{\hbox{/\kern-.5600em$\epsilon$}}
\def\delsl{\hbox{/\kern-.7000em$\nabla$}}
\def\lxpsl{\hbox{/\kern-.5600em$l$}}
\def\kxpsl{\hbox{/\kern-.5600em$k$}}
\def\qxpsl{\hbox{/\kern-.5600em$q$}}
\def\sla#1{\raise.15ex\hbox{$/$}\kern-.57em #1}
\def\pwr#1{\cdot 10^{#1}}



\def\roughly#1{\mathrel{\raise.3ex\hbox{$#1$\kern-.75em\lower1ex\hbox{$\sim$}}}}

\def\ol#1{\overline{#1}}



\def\bfp{{\bf p}}



\def\Scl{{\cal L}}
\def\Scm{{\cal M}}


\def\Im{{\rm Im\;}}
\def\diag#1{{\rm diag}\left( #1 \right)}





\def\hc{{\rm h.c.}}

\def\edm{e.d.m.}


\def\GeV{{\rm \ GeV}}

\def\ecm{{\it e}{\hbox{\rm -cm}}}



\ref\gim{S.~Chivukula and H.~Georgi, \plb{188}{87}{99}.}

\ref\cantdo{Our experimental colleagues tell us that detection
of the photon polarization in these decays may be very difficult.}

\ref\loggim{M.A.~Shifman, A.I.~Vainshtein and V.I.~Zakharov,
\prd{18}{78}{2583}.}

\ref\edmbound{I.S. Altarev \etal, \jetp{44}{86}{460}; 
K.F. Smith \etal, \plb{234}{90}{191}.}


\ref\hyperonth{F.J.~Gilman and M.B.~Wise, \prd{19}{79}{976};
S.G.~Kamath, \npb{198}{82}{61};
L.~Chong-Huah, \prd{26}{82}{199};
J.O.~Eeg, \zpc{21}{84}{253};
M.K.~Gaillard and X.Q.~Li, \plb{158}{85}{158}.}

\ref\omegaref{L. Bergstr\"om and P. Singer, \plb{169}{86}{297}.}

\ref\pdbk{Particle Data Group Review of Particle Properties, \plb{239}{90}{}.}

\ref\lowerbound{Ya.I.~Kogan and M.A.~Shifman,
\sjnp{38}{83}{1045}{38}{83}{628}.}

\ref\bdecayth{S.~Bertolini, F.~Borzumati and A.~Masiero, \prl{59}{87}{180};
N.G.~Deshpande, P.~lo, J.~Trampetic, G.~Eilam and P.~Singer, \prl{59}{87}{183};
R.~Grigjanis, P.~O'Donnell, M.~Sutherland and H.~Navelet, \plb{213}{88}{355};
B.~Grinstein, R.~Springer and M.~Wise, \plb{202}{88}{138}.}



\def\fcnc{flavour-changing neutral currents}

\def\chiral{$SU_L(3) \times SU_{u_R}(3) \times SU_{d_R}(3)$}
\def\uyuk{\lambda^u}
\def\dyuk{\lambda^d}

\def\ddsyuk{\lambda^d \lambda^{d\dagger}}

\def\suuyuk{\lambda^{u\dagger} \lambda^u}
\def\sddyuk{\lambda^{d\dagger} \lambda^d}


\rightline{April 1992}
\rightline{McGill-92/18}
\rightline{BNL-47586}
\vskip .2in

\title
\centerline{A General Lower Bound for}
\centerline{Radiative B Decay}
\endtitle

\authors
\centerline{David Atwood${}^a$, C.P. Burgess${}^b$ and A.
Soni${}^a$\footnote{}{email: atwood@bnlcl1.bnl.gov; cliff@physics.mcgill.ca;
soni@bnlcl1.bnl.gov}}
\vskip .15in
\centerline{\it ${}^a$ Department of Physics,
Brookhaven National Laboratory}  \centerline{\it Upton New York 11973 USA.}
\vskip .1in
\centerline{\it ${}^b$ Physics Department, McGill University}
\centerline{\it 3600 University St., Montr\'eal, Qu\'ebec, CANADA, H3A 2T8.}
\endauthors

\abstract
In a wide class of models -- including the standard model --
\fcnc\ are suppressed by an approximate chiral symmetry of the
low-energy lagrangian. This symmetry allows the derivation of
general relations among low-energy flavour-changing processes.
We derive one such: the relative sizes of the decay rates $b \to
s \gamma$, $b \to d \gamma$ and $s \to d \gamma$. Together with a
unitarity-related  {\it lower} bound on the rate for the strangeness-changing
hyperon decay $\Omega^- \to \Xi^- \gamma$ we obtain a reasonably
model-independent {\it lower} bound for the inclusive $b \to s \gamma$ rate.
\endabstract


\section{Introduction}

Flavour-changing neutral-current processes have long been known to
be extremely rare. They are much weaker than would be naively expected to be
generated through loops involving the known flavour-changing charged-current
interactions. This small size is understood within the standard model as being
due to the GIM mechanism: a partial cancellation among loop-induced effects.

This fairly delicate cancellation provides a major clue as to the nature of
the new physics that is expected to replace the standard model at higher
energies. A minimal requirement for any candidate model for such new physics
is that it not ruin this cancellation and so generate unacceptably large
\fcnc.

In the absence of any direct knowledge of the physics at higher scales, and
barring the possibility of an unnatural fine tuning, the absence of such
flavour-changing effects is most efficiently summarized in terms of an
approximate symmetry of the low-energy effective lagrangian that is obtained
once the unknown heavy particles are integrated out of the underlying theory.
Such a symmetry formulation has arisen \gim\ within the context of technicolor
models, for example.

The main purpose of this note is to point out that such an approximate symmetry
strongly constrains the generation dependence of {\it all} of the interactions
in such an effective theory. As such it leads to general relations amongst the
potential flavour-changing effects that can be probed purely at low energies.
Most importantly, since these relations follow purely from the symmetries of
the
low-energy effective theory their validity transcends that of any particular
model for new physics and may be considered a general consequence of the given
symmetry-breaking pattern.

We illustrate this point here by using such an approximate symmetry to derive
the relative strength of the flavour-changing processes $s \to d \gamma$,
$b \to s \gamma$ and $b\to d \gamma$. This relation holds in a very broad class
of models, including the standard model. Using a {\it lower} bound for the $s
\to d \gamma$ rate which follows from unitarity we are then able to derive a
{\it
lower} bound for the inclusive branching ratio $b \to s \gamma$ which turns out
to be 8.5 times smaller than the standard model prediction. For $100 \,\GeV <
m_t < 200 \,\GeV$ the lower bound is $B(b \to s \gamma) > (3 - 5) \pwr{-5}$.
The
corresponding inclusive $b \to d \gamma$ bound is twenty-five times smaller:
$B(b
\to d \gamma) > (1 - 2) \pwr{-6}$. Should these decays not be measured at this
level then this would serve to rule out not just the standard model but any
model
within which \fcnc\ are suppressed \via\ the same type of approximate symmetry.

The weakest link in the arguments that establish this lower bound is
the extraction of the quark-level bound for $s\to d \gamma$ from the lower
limit on the rate for radiative $\Omega^-$ decay. This relies on the expected
dominance in this decay of the single-quark transition. An experimental check
on
this reasoning is given by polarization-sensitive measurements such as the
decay
distribution of outgoing photons relative to the initial $\Omega^-$
polarization. The single-quark picture of this decay unambiguously predicts an
asymmetry parameter $\alpha = -1$. A similar check for $b$ decays involves
detecting the correlation between the direction of the final hadrons relative
to
the photon polarization in decays such as
$B \to K^*\gamma \to K \pi \gamma$. \cantdo\

We now turn to the guts of our argument. We start by briefly summarizing the
formulation of the GIM mechanism in terms of an approximate symmetry. This is
next used to fix the relative size of the quark-level amplitudes for $b\to s
\gamma$, $b\to d \gamma$ and $s \to d \gamma$ leading to the lower bound on the
total branching ratio for radiative $b$ decay. We conclude with a short
discussion of two possible polarization-dependent observables which might be
used to check the validity of the single-quark mechanism for mediating these
flavour-changing transitions once they have been reliably observed.

\section{The Approximate Symmetry}

The standard model enjoys an approximate \chiral\ symmetry under which the
quark
fields transform in the following representation
\eq
\label\chtransfn
\pmatrix{ u_i \cr d_i \cr}_L  \sim ({\bf 3}, {\bf 1}, {\bf 1}), \quad
(u_a)_R \sim ({\bf 1}, {\bf 3}, {\bf 1}), \quad
(d_n)_R \sim ({\bf 1}, {\bf 1}, {\bf 3}).
\eeq
If unbroken, this symmetry would completely forbid \fcnc. In the standard model
this symmetry is broken only by the Yukawa couplings which generate fermion
masses. As a result the generation dependence of any flavour-changing
neutral-current amplitude is completely determined by its dependence on these
Yukawa couplings. Once expressed in a basis of mass eigenstates for which the
Yukawa couplings are diagonal this ensures the GIM-type suppression of all
flavour changing  amplitudes by fermion masses and Kobayashi-Maskawa
mixing angles.

Once phrased in this manner it is clear how to ensure the same properties for a
general low-energy effective lagrangian which need {\it not} include scalars
and
Yukawa couplings. The effective lagrangian must first enjoy the same
approximate
\chiral\ symmetry with the fermion transformation law of eq.~\chtransfn. The
content of the GIM mechanism may then be summarized by the requirement that the
unknown underlying physics only involve two types of (small) order parameters
which break this symmetry by `transforming' in the following way
\eq
\label\yuktransfn
\uyuk \sim (\ol{\bf 3},{\bf 3},{\bf 1}), \quad
\dyuk \sim (\ol{\bf 3},{\bf 1},{\bf 3}).
\eeq
Notice that these transformation rules are precisely those that are
satisfied by the Yukawa couplings in the standard model.

This transformation property ensures an acceptable quark mass matrix:
\eq
\label\massterm
\Scl_m = - \frac{M}{2} \left[ \uyuk_{ai} \; \ol{u}_a \, \gamma_L \, u_i +
\dyuk_{ni} \; \ol{d}_n \, \gamma_L \, d_i \right] + \hc,
\eeq
in which $M$ is a large new-physics scale which might reasonably be several
hundreds of GeV. Comparison with the known masses would then imply that $\uyuk$
and $\dyuk$ have the same flavour-structure and size as have the standard-model
Yukawa couplings.


Consider now the implications that may be inferred from such an approximate
symmetry for radiative flavour-changing transitions. Lorentz invariance
requires that the matrix element for the process $d_i \to d_n \gamma$ must take
the general form
\eq
\label\dimfive
\Scm(d_i \to d_n \gamma) = \frac{e}{M} \, \xi_{ni} \; \ol{d}_n \,
\gamma^{\mu\nu} \gamma_L \, d_i \; q_\nu \, \epsilon_\mu(q),
\eeq
where $q$ and $\epsilon_\mu(q)$ are the photon's four-momentum and
polarization.

The main point is now this: for small $\uyuk$ and $\dyuk$ the generation
structure of the coefficients $\xi_{im}$ is determined up to an overall
normalization. For example, since the operator in eq.~\dimfive\ transforms
under
the chiral symmetry as $({\bf 3},{\bf 1},\ol{\bf 3})$ we must have
\eq
\label\genform
\xi_{ni} =  \dyuk_{nj} \; \left[ A(\suuyuk,\sddyuk) \right]_{ji} + \left[
B(\ddsyuk) \right]_{nm} \; \dyuk_{mi}.  \eeq
Should these functions be analytic near $\uyuk= \dyuk =0$, $A$ and $B$
could be replaced by their Taylor expansions about this point. In the standard
model, however, these functions turn out to be logarithmically singular for
small
$\uyuk$ and $\dyuk$ \loggim\ once QCD corrections are included. This leads
instead to  the following asymptotic expression
\eqa
\label\running
A &\simeq a_1 \; \ln\left(\suuyuk \right) + a_2 \; \ln\left(\sddyuk \right)
+ a_3 + a_4 \; \suuyuk + a_5 \sddyuk + \cdots ,
\eol
\hbox{and}\qquad B &\simeq b_1 \; \ln\left(\ddsyuk \right) +
b_2 + b_3 \ddsyuk + \cdots . \eeol
\eeq

The GIM mechanism now reveals itself once these interactions are expressed in a
basis of fermion mass eigenstates: $u_{Li} = (U_u)_{ij} \; u'_{Lj}$,
$u_{Ra} = (U_u)^*_{ab} \; u'_{Rb}$, $d_{Li} = (U_d)_{ij} \; d'_{Lj}$ and
$d_{Rm} = (U_d)^*_{mn} \; d'_{Rn}$. The unitary matrices $U_u$ and $U_d$ here
are chosen to diagonalize the appropriate mass matrices: $m^u = U_u^T
\uyuk U_u \; M = \diag{m_u,m_c,m_t}$ and $m^d = U_d^T \dyuk U_d \; M
= \diag{m_d,m_s,m_b}$. In this basis all terms in eq.~\genform\ that
involve only $\dyuk$ become automatically diagonal. As a result the only
off-diagonal coefficient which can remain in eq.~\dimfive\ is
\eq
\label\massbasis
\xi_{ni} =  \sum_{x=u,c,t} V^*_{xn} V_{xi} \; \frac{m^d_n}{M} \left\{
a_1 \;  \ln\left[\frac{(m^u_x)^2}{M^2} \right] + a_4
\frac{(m^u_x)^2}{M^2} + \cdots \right\}.
\eeq
The matrix $V = U_u^\dagger U_d$ that appears in this expression is the usual
Kobayashi-Maskawa matrix whose elements govern the charged-current weak
interactions.

In a world for which the QCD scale is much smaller than the current
quark masses the logarithmic dependence on $\uyuk$ and $\dyuk$ reflects an
infrared mass singularity as the up-quark masses tend to zero. In the real
world, for which $\Lambda_{QCD} > m_u$, such a dependence on $m_u$ requires
some comment. In this case, although the $m_c$ and $m_t$-dependence do not
change, the simple perturbative expression for the $m_u$-dependence gets
complicated by perturbatively incalculable QCD effects. The logarithmic
dependence on $m_c$ and $m_t$, together with the chiral \chiral\ symmetry
nevertheless still imply logarithmic dependence of $A$ and $B$ on $\uyuk$. This
must then combine with strong QCD contributions to the nonlogarithmic terms ---
such as $a_3$ --- to convert $\ln(m_u^2/M^2)$ into a logarithm involving an
appropriate hadronic scale with the same chiral behaviour: $\ln(m_\pi^4/M^4)$.

As advertised, to the extent that powers of $\uyuk$ may be neglected, only the
single complex number $a_1$ in eq.~\massbasis\ is left undetermined by the
approximate chiral symmetry. In this limit this is the only quantity which can
distinguish various underlying theories from one another.

Notice also that the
explicit dependence of the coefficients of this equation on the quark masses
and
on the Kobayashi-Maskawa matrix precisely mirrors those that are obtained when
this vertex is generated by loops within the standard model.
In the standard-model case $a_1$ turns out to be real. More generally a
nontrivial phase for $a$ is strongly bounded by limits on the neutron electric
dipole moment. This is because such a phase violates $CP$ and so generates a
quark electric dipole moment
\eq
\label\edm
D_f =  \Im  {\it a}  \; \frac{m_f}{M^2} \; \sum_{x=u,c,t} |V^*_{xf}|^2
\ln\left[
\frac{(m^u_x)^2}{M^2} \right],
\eeq
which must be less than $\sim 10^{-26} \, \ecm$ \edmbound.


\section{A Lower Bound for $b\to s \gamma$}

As is clear from the previous section, the approximate \chiral\ symmetry
relates the strength of the effective $s \to d$, $b \to d$ and $b \to s$
operators in the following way
\eqa
\label\relation
\frac{\Gamma(b\to s\gamma)}{\Gamma(s\to d \gamma)} &= \frac{m_b^2}{m_s^2} \;
\left[ \frac{ V^*_{cs} V_{cb} \ln\left( \frac{m_c^2}{m_u^2} \right) + V^*_{ts}
V_{tb} \ln\left( \frac{m_t^2}{m_u^2} \right) }{ V^*_{cd} V_{cs} \ln\left(
\frac{m_c^2}{m_u^2} \right) + V^*_{td} V_{ts} \ln\left( \frac{m_t^2}{m_u^2}
\right) }\right]^2. \eolnn
&\simeq (10 - 20) \qquad \hbox{for} \qquad 100 \, \GeV < m_t < 200 \, \GeV;
\eol
\frac{\Gamma(b\to d\gamma)}{\Gamma(b\to s \gamma)} &=
\left[ \frac{ V^*_{cd} V_{cb} \ln\left( \frac{m_c^2}{m_u^2} \right) + V^*_{td}
V_{tb} \ln\left( \frac{m_t^2}{m_u^2} \right) }{ V^*_{cs} V_{cb} \ln\left(
\frac{m_c^2}{m_u^2} \right) + V^*_{ts} V_{tb} \ln\left( \frac{m_t^2}{m_u^2}
\right) }\right]^2. \eolnn
&\simeq 0.04 \qquad \hbox{for} \qquad 100 \, \GeV < m_t < 200 \, \GeV. \eeol
\eeq
We use current-quark masses in the numerical estimates and take $m_s \simeq
200$ MeV.

We next wish to use this relation together with experimental information
concerning the process $s\to d$ to derive a lower bound for the inclusive
$b \to s \gamma$ and $b \to d \gamma$ rates.  Our starting point is the
recognition
that the radiative $s \to d$ transition can be observed in radiative hyperon
decays. Although detailed study of these decays over the past ten years
\hyperonth\ has indicated that they are generally {\it not} dominated by the
single-quark decay $s \to d \gamma$, there are exceptions to this rule. In
particular, the single-quark contribution appears to dominate the decay
$\Omega^- \to \Xi^- \gamma$ -- dominating both penguin and long-distance
effects
\omegaref. An evaluation of the matrix elements in
ref.~\omegaref\ gives the result
\eq
\label\smresult
B_{\rm sm}(\Omega^- \to \Xi^- \gamma) \simeq  6.8 \pwr{-5} \qquad \hbox{for}
\qquad \xi_{ds} = 0.8 \; \frac{G m_s}{16 \pi^2}.
\eeq
$G$ here represents the usual Fermi constant.

Unfortunately this decay has not yet been observed, the present 90\% c.l. upper
bound being $B_{\rm exp}(\Omega^- \to \Xi^- \gamma) < 2.2 \pwr{-3}$ \pdbk.
Experiment does, however, furnish a {\it lower} bound to this branching ratio
\lowerbound\ as may be seen by cutting the long-distance contributions such as
that shown in Fig. 1. Both halves of the cut diagram may be determined from
experiment together with $SU_f(3)$ flavour symmetry. The lower bound obtained
in
this way in ref.~\lowerbound\ is $B(\Omega^- \to \Xi^- \gamma) > 8\pwr{-6}$.
When taken together with the dominance of the single-quark decay in this mode,
this gives the following lower bound
\eq
\label\sdlowerbound
\frac{B(b\to s \gamma)}{B_{\rm sm}(b \to s \gamma)} = \frac{B(s\to d
\gamma)}{B_{\rm sm}(s \to d \gamma)} = \frac{B(\Omega^- \to \Xi^- \gamma)}{
B_{\rm sm}(\Omega^- \to \Xi^- \gamma)} > 0.12.
\eeq
Together with the standard-model prediction \bdecayth: $B_{\rm sm}(b\to
s\gamma) = (2 - 4)\pwr{-4}$ for $50 \, \GeV < m_t < 200 \, \GeV$ this leads to
the lower bounds quoted above:
\eq
\label\bsbound
B(b \to s \gamma) > (3 - 5) \pwr{-5}, \qquad \hbox{and} \qquad B(b \to d
\gamma)
> (1 - 2) \pwr{-6}.
\eeq

This establishes the main result of this letter. We now turn to a brief
discussion of how the measurement of polarization-sensitive observables in
these decays can be used to check the importance of the single-quark
mechanism in both hyperon and $B$ decays.

\section{Polarized Observables}

Until flavour-changing radiative $b$ decays are directly observed it is
necessary to look to strange-quark processes in order to determine the
normalization of the interaction of eq.~\dimfive. As outlined in the previous
section we have used for these purposes the rate for the hyperon decay
$\Omega^-
\to \Xi^- \gamma$. The key part of the argument is the expectation that it is
the single-quark process that is responsible for this particular decay. In
ref.~\omegaref\ this conclusion is reached by comparing the single-quark rate
with that arising from long-distance contributions as well as penguin-type
$s\to d g$ gluon transitions.

Much of the uncertainty in these arguments may be removed once the $\Omega^-
\to \Xi^- \gamma$ decay is directly observed. There are two reasons for this.
The first reason is that once the rate for this decay is experimentally known
it
can be compared to the total inclusive rates or upper limits for $b\to s\gamma$
and $b\to d \gamma$. More importantly, once this hyperon decay is observed it
becomes possible to measure the decay distribution of the $\Xi^-$ relative to
the direction of $\Omega^-$ polarization in its rest frame.

As a function of the mass ratio, $x\equiv m_\Xi/m_\Omega$, and the cosine,
$z=\cos\theta$, of the angle in the $\Omega^-$ rest frame between
the $\Xi^-$ momentum and the
direction of $\Omega^-$ polarization the decay distribution is given by:
\eqa
\label\distribution
\left[ \nth{\Gamma} \; \frac{d\Gamma}{dz} \right]_{h=3/2} &=
\frac{3}{4(x^2+ 3)} \Bigl[(1+x^2) + (3-x^2) z^2 - \alpha z \Bigl( (3+x^2) -
(1-x^2) z^2 \Bigr) \Bigr], \eol
\left[ \nth{\Gamma} \; \frac{d\Gamma}{dz} \right]_{h=1/2} &=
\frac{1}{4(x^2+3)} \Bigl[ (x^2+9) + (3x^2-9) z^2 + \alpha z \Bigl( (5x^2-9) +
9(1-x^2) z^2 \Bigr) \Bigr]. \eeol
\eeq
The shape parameter $\alpha$ in this equation is predicted to satisfy
$\alpha=-1$ if the decay is dominantly mediated by the single-quark
interaction of eq.~\dimfive. A measurement of this value for $\alpha$ would
provide convincing support for the single-quark mechanism for the decay.

A similar check would also be useful when diagnosing flavour-changing radiative
$B$-meson decays. This is particularly so for the $b\to d$ transition for which
the single-quark interpretation is less clean than it is for $b\to s$.
Unfortunately such observables are harder to come by for $B$ mesons.
One possibility arises should the polarization of the outgoing photon prove
to be measureable \cantdo\. In this case the parity-preserving
and parity-violating
contributions to the $b \to s$ or $b\to d$ transition operator of eq.~\dimfive\
may be disentangled from one another by measuring the angular distribution
of decay products relative to the direction of
polarization of the photon.

For example, in the decay $B \to K^* \gamma \to K \pi \gamma$ (or $B \to \rho
\gamma\to \pi \pi \gamma$) choose, in the $B$ rest frame, the directions of
the momentum and polarization of the outgoing photon to define the $z$- and
$x$-axes. Denote in this frame the polar angles defining the direction
of the momentum difference $\bfp_\pi - \bfp_K$ (or $\bfp_{\pi_1} -
\bfp_{\pi_2}$)
by $(\theta,\phi)$. Then the decay distribution as a function of $\phi$
is predicted to behave as
\eq
\label\bdist
\nth{\Gamma} \; \frac{d\Gamma}{d\phi} = \frac{1}{2\pi} \left[ 1 + \sqrt{1 -
\alpha^2} \; \cos 2\phi \right].
\eeq
As before $\alpha = -1$ if the single-quark transition dominates the decay.

\section{Conclusions}

In summary, we have shown how the natural expression of the observed
suppression
of \fcnc\ in terms of an approximate chiral symmetry of the effective
lagrangian leads to definite relations amongst low-energy flavour-changing
processes. We show in particular that within a wide class of models --
including, of course, the standard model -- the relative strength of the
quark-level radiative transitions $s\to d\gamma$, $b\to s\gamma$ and $b\to d
\gamma$ is completely fixed in terms of particle masses and Kobayashi-Maskawa
mixing angles.

Although this same prediction is well known within the standard-model context,
our derivation makes explicit the much wider validity  of the result. This
would be of most interest should these predictions fail, in which case a broad
framework for suppressing \fcnc\ would be ruled out.

We finally combine the relation among these amplitudes with two facts
concerning
the hyperon radiative decay $\Omega^- \to \Xi^- \gamma$ in order to establish a
model-independent {\it lower} bound for the inclusive rates $B(b\to s\gamma)$
and $B(b\to d\gamma)$. The two facts needed to obtain these results are ($i$)
the long-distance unitarity lower bound for $B(\Omega^- \to \Xi^-\gamma)$
derived
in ref.~\lowerbound, and ($ii$) the dominance for this process of the
single-quark decay found in ref.~\omegaref.

The lower bounds obtained in this way are $B(b\to s \gamma) > (3 - 5)\pwr{-5}$
and $B(b\to d\gamma) > (1 - 2)\pwr{-6}$. The ranges quoted indicate the
dependence of the result on the top mass which we take to lay between 100 and
200 GeV. Of these bounds the first is of most immediate interest given that
branching ratios of the order of $10^{-4}$ are presently being probed at CLEO.

\bigskip
\centerline{\bf Acknowledgments}
\bigskip

C.B. would like to thank Markus Luty, Drew Peterson and Ken Ragan for various
flavour-changing discussions. This research was partially funded by funds from
the N.S.E.R.C.\ of Canada and les Fonds F.C.A.R.\ du Qu\'ebec. The research of
D.A. and A.S. has been partially supported under contract number
DE-AC02-76CH00016 with the US Department of Energy.

\bigskip
\centerline{\bf Figure Caption}
\bigskip

A long-distance contribution to the decay $\Omega^- \to \Xi^-\gamma$ from
whose imaginary part the lower bound is derived.

\listrefs

\bye